\documentstyle[prc,aps,manuscript]{revtex}

\begin{document}

\draft

\title{LOW DENSITY INSTABILITY IN A NUCLEAR FERMI LIQUID DROP}

\author{V.M. Kolomietz$^{1,2)}$ and S. Shlomo$^{1)}$}

\address{$^{1)}$Institute for Nuclear Research,
Prosp. Nauki 47, 252028 Kiev, Ukraine}
\address{$^{2)}$Cyclotron Institute, Texas A\&M University,
College Station, Texas 77843}


\maketitle

\begin{abstract}
The instability of a Fermi-liquid drop with respect to bulk density 
distortions is considered. It is shown that the presence of the 
surface strongly reduces the growth rate of the bulk instability of 
the finite Fermi-liquid drop because of the anomalous dispersion 
term in the dispersion relation. The instability growth rate is 
reduced due to the Fermi surface distortions and the relaxation 
processes. The dependence of the bulk instability on the multipolarity
of the particle density fluctuations is demonstrated for two nuclei 
$^{40}Ca$ and $^{208}Pb$.
\end{abstract}

\vskip 1cm


\section{Introduction}
In the vicinity of the equilibrium state of the nuclear Fermi-liquid
drop the stiffness coefficients are positive and the system is
stable with respect to particle density and surface distortions. With 
decreasing bulk density or increasing internal excitation energy
(temperature) the liquid drop reaches the regions of  mechanical
or thermodynamical instabilities with respect to small particle
density and shape fluctuations and to separation into liquid and
gas phases. The process of the development of instability is a
complicated one. We will discuss some aspects of the process. 
In particular, we will study the influence of a sharp liquid-drop 
boundary on the instability with respect to small particle density 
fluctuations. In actual nuclear processes (heavy-ion reactions, nuclear 
fission etc.) nuclear matter is not static, and consequently the 
development of instability depends not only on the equation of state, 
but also on the dynamical effects such as the dynamical Fermi-surface
distortion or the relaxation processes. We will take into account 
these aspects in studying the stability of the Fermi-liquid drop
in both regimes of the first- and zero sound modes.

\section{Bulk instability of the Fermi-liquid drop}
Let us consider small density fluctuations $\delta \rho ({\bf r}, t)$
starting from the  nuclear fluid dynamic approach \cite{RiScb,book}. 
The linearized equation of motion reads (see Ref. \cite{KiKoSh}),
\begin{equation}
m\,{\partial^2\over \partial t^2}\,\delta \rho =
\vec{\nabla}\rho_{eq}\,\vec{\nabla}{\delta E\over \delta \rho} +
\nabla_\nu\nabla_\mu P_{\nu \mu}^\prime,
\label{eq1}
\end{equation}
where $\rho_{eq}$ is the equilibrium density, $E$ is the total energy
and the pressure tensor, $P_{\nu \mu}^\prime$, represents the deviation 
of the pressure from its isotropic part due to the Fermi surface 
distortions.

The variational derivative $\delta E/\delta \rho$ in Eq. (\ref{eq1}) 
implies a linearization with respect to the density variation 
$\delta \rho$:
\begin{equation}
{\delta E\over \delta \rho} = \Big({\delta E\over \delta \rho}
\Big)_{eq} + \hat{L}[\rho_{eq}]\,\delta \rho + {\cal O}\Big(\delta \rho
\Big)^2.
\label{de1}
\end{equation}
We point out that the first term on the r.h.s. of Eq. (\ref{de1})
does not enter Eq. (\ref{eq1}) because of the the equilibrium condition 
$(\delta E/\delta \rho)_{eq} = \lambda_F = {\rm const}$, where 
$\lambda_F$ is the chemical potential. The operator $\hat{L}$ can be 
derived from the equation of state $E = E[\rho ]$. We will use the 
extended Thomas-Fermi approximation for the internal kinetic energy 
\cite{BhRo} and the Skyrme-type forces for the interparticle interaction 
\cite{EnBrGo}. In the special case of a spin saturated and charge 
conjugated nucleus, and neglecting spin-orbit and Coulomb effects,
the equation of state reads
$$
E[\rho ] = \int d{\bf r} \,\,\Big\{{\hbar^2\over 2\,m}\,
\left[{3\over 5}\,\Big({3\pi^2\over 2}\Big)^{2/3} \,\rho^{5/3} +
{1\over 4}\,\eta \,{(\vec{\nabla}\rho)^2\over \rho}\right]
$$
\begin{equation}
+ {3\over 8}\,t_0\,\rho^2 + {1\over 16}\,t_3\,\rho^3 +
{1\over 64}\,(9\,t_1 - 5\,t_2) \,(\vec{\nabla}\rho)^2\Big\}.
\label{e1}
\end{equation}
The effective forces used in Eq. (\ref{e1}) leads to an overestimate
of the incompressibility coefficient. This is a well-known feature of 
Skyrme forces which can be overcome by taking non-integer powers of 
$\rho$ in the potential energy density in Eq. (\ref{e1}). For our 
purposes we shall, however, be content with the form (\ref{e1}). To 
make quantitative estimates of the finite size effects on the bulk 
instability of the liquid drop, we will assume a sharp surface 
behaviour of $\rho_{eq}({\bf r})$ having a bulk density $\rho_0$ and 
an equilibrium radius $R_0$. Taking into account Eqs. (\ref{de1}) and 
(\ref{e1}), the operator $\hat{L}[\rho_{eq}]$ is then reduced to the 
following form
\begin{equation}
\hat{L}[\rho_{eq}]\,\delta\rho = 
{K\over 9}\,\nabla^2\,\delta\rho - 2\,(\beta + t_s\,\rho_0)\,
\nabla^2\,\nabla^2\,\delta\rho\,\,\,\,\,\, {\rm at}\,\,\,\,r < R_0,
\label{hatl}
\end{equation}
where
$$
\beta = {\hbar^2\over 8\,m}\,\eta,\,\,\,\,\,\,
t_s = {1\over 64}\,(9\,t_1 - 5\,t_2)
$$
and $K$ is the incompressibility coefficient
\begin{equation}
K = 6\,e_F\,(1 + F_0)\,\left(1 + {1\over 3}\,F_1\right)^{-1}.
\label{K1}
\end{equation}
The Landau parameters $F_l$ are given by
\begin{equation}
F_0 = {9\,\rho_0\over 8\,\epsilon_F}\,
\left[t_0 + {3\over 2}\,t_3\,\rho_0\right]\,{m^\ast\over m}
+ 3\,\left(1 - {m^\ast\over m}\right),\,\,\,
F_1 = 3\,\left({m^\ast\over m} - 1\right),
\label{Fl}
\end{equation}
where
$$
{m\over m^\ast} = 1 + {m\,\rho_0\over 8\,\hbar^2}\, 
(3\,t_1 + 5\,t_2)
$$
and $e_F$ is the Fermi energy.

The pressure tensor $P_{\nu \mu}^\prime$ can be expressed through the 
displacement field $\vec{\chi}({\bf r}, t)$ \cite{KoMaPl}. Assuming 
also $\delta \rho \sim e^{-i\omega t}$, the pressure tensor 
$P_{\nu \mu}^\prime$ is given by \cite{KiKoSh}
\begin{equation}
P_{\nu\mu}^\prime = {i\omega \tau  
\over {1 - i\omega \tau}} P_{eq} \Lambda_{\nu\mu}, 
\label{p2}
\end{equation}
where $\tau$ is the relaxation time and we used the symbol
\begin{equation}
\Lambda_{\nu\mu} = \nabla_\nu \chi_{\mu} +
\nabla_\mu \chi_{\nu} - {2 \over 3}\delta_{\nu\mu}\nabla_
{\lambda}\chi_{\lambda} 
\label{lambda}
\end{equation}
for this combination of gradients of the Fourier transform $\chi_{\nu}$ 
of the displacement field $\vec{\chi}({\bf r}, t)$. The equilibrium 
pressure of a Fermi gas, $P_{eq}$, in Eq. (\ref{p2}), is given by
$$
P_{eq} = {1\over 3m}\int {d{\bf p}\over (2\pi \hbar)^3}\,\,
p^2\,f_{eq}({\bf r}, {\bf p}) \approx \rho_0\,p_F^2/5\,m,
$$
where $f_{eq}({\bf r}, {\bf p})$ is the equilibrium phase-space
distribution function and $p_F$ is the Fermi momentum. We point out 
that Eq. (\ref{p2}) is valid for arbitrary relaxation time $\tau$ and 
thus describes both the zero- and the first-sound limit as well as the 
intermediate case.

Taking into account the continuity equation and Eqs. (\ref{hatl}), 
(\ref{p2}) and (\ref{lambda}), the equation of motion (\ref{eq1}) can 
be reduced in the nuclear interior to the following form (we consider 
the isoscalar mode):
\begin{equation}
- m\,\omega^2 \,\delta\rho = \left({1\over 9}\,K -
{4\over 3}{i\omega \tau  
\over {1 - i\omega \tau}} (P_{eq}/\rho_0)\right)
\nabla^2\delta \rho 
- 2\,(\beta + t_s\,\rho_0)\,\nabla^2 \nabla^2 \delta\rho.
\label{0eq3}
\end{equation}

The solution of Eq. (\ref{0eq3}) for a fixed multipolarity $L$ is 
given by
\begin{equation}
\delta \rho ({\bf r}, t) = \rho_0 \,j_L (qr)\,Y_{LM}(\theta, \phi)\,
\alpha_{LM}(t),
\label{drhoL}
\end{equation}
where $q$ is the wave number and $\alpha_{LM}(t)$ is the amplitude of 
the density oscillations. We will distinguish between stable and 
unstable regimes of density fluctuations. In the case of a stable mode 
at $K > 0$, a solution of Eq. (\ref{0eq3}) of the form (\ref{drhoL}) 
has the following dispersion relation
\begin{equation}
\omega^2 = u^2\,q^2 - i\,\omega\,
{\gamma (\omega )\over m}\,q^2 + \kappa_s \,q^4.
\label{disp1}
\end{equation}
Here, $u$ is the sound velocity
\begin{equation}
u^2 = u_1^2 + \kappa_v,
\label{c}
\end{equation}
where $u_1$ is the velocity of the first sound
\begin{equation}
u_1^2 = {1\over 9\,m}\,K, 
\label{c1}
\end{equation}
$\gamma (\omega )$ is the viscosity coefficient
\begin{equation}
\gamma (\omega ) = {4\over 3}{\rm Re} \Big({\tau  
\over {1 - i\,\omega \,\tau }}\Big) {P_{eq}\over \rho_{eq}}
\label{gamma}
\end{equation}
and
\begin{equation}
\kappa_v = {4\over 3}{\rm Im} \Big({\omega \tau
\over {1 - i\,\omega \,\tau }}\Big) {P_{eq}\over m\,\rho_{eq}},\,\,\,\,
\,\,\,\,\,\,\,\kappa_s = {2\over m}\,(\beta + t_s\,\rho_0).
\label{kappa}
\end{equation}
The quantities $\kappa_v$ and $\gamma (\omega )$ appear due to the 
Fermi-surface distortion effect. The dispersion relation (\ref{disp1}) 
determines both the real and the imaginary part of the eigenfrequency 
$\omega$.

The equation of motion (\ref{0eq3}) has to be augmented by the boundary
condition. This is given by a condition of the balance of the surface 
pressure $\delta P_{surf}$ with the volume sound pressure 
$\delta P_{sound}$ on a free surface of the liquid drop, see Refs.
\cite{Lamb,BoMo}. It reads
\begin{equation}
m\,u^2\,\rho_0\,j_L(qR_0) = {1\over q^2\,R_0^2}\,
(L-1)\,(L+2)\,\sigma\,{\partial j_L(qr)\over
\partial r}\Big\vert_{r=R_0}.
\label{sec1}
\end{equation}

Let us consider now the volume instability regime, $K < 0$, and
introduce a growth rate $\Gamma = -i\,\omega$ ($\Gamma$ is real,
$\Gamma > 0$), see Ref. \cite{PeRa1}. Using Eq. (\ref{disp1}), one 
obtains
\begin{equation}
\Gamma^2 = |u_1|^2 \,q^2 - \zeta (\Gamma )\,q^2 - \kappa_s\,q^4,
\label{0disp}
\end{equation}
where
\begin{equation}
\zeta (\Gamma ) = {4\over 3\,m}{\Gamma \tau \over
{1 + \Gamma \tau}}{P_{eq}\over \rho_0}.
\label{zeta}
\end{equation}

Equation (\ref{0disp}) is valid for arbitrary relaxation time $\tau$.
From it one can obtain the leading order terms in the different limits 
mentioned above. 

{\it (i) Frequent collision regime: $\tau \to 0$.}\\
The contribution from the dynamic distortion of the Fermi surface,
$\kappa_v$, can be neglected in this case and we have from Eqs.
(\ref{0disp}) and (\ref{gamma}),
\begin{equation}
\Gamma^2 = |u_1|^2 \,q^2 - \Gamma\,(\tilde{\gamma}/m)\,q^2 - \kappa_s\,q^4,
\label{0disp1}
\end{equation}
where $\tilde{\gamma}=(8/15)\,e_F\,\tau$ is the viscosity
coefficient.
In the case of small viscosity coefficient $\tilde{\gamma}$, one
has from Eq. (\ref{0disp1})
\begin{equation}
\Gamma^2 \approx |u_1|^2\,q^2 - \kappa_s\,q^4 -
{\tilde{\gamma}\over m}\,q^2\,\sqrt{|u_1|^2\,q^2 - \kappa_s\,q^4}.
\label{0disp2}
\end{equation}

The amplitude of the density oscillations, $\delta \rho_L({\bf r}, t)$,
grows exponentially if $\Gamma > 0$. Expression (\ref{0disp2})
determines two characteristic values of the wave number $q$, namely,
$q_{max}$ where the growth rate reaches a maximum of $\Gamma_{max}$, and 
$q_{crit}$ where $\Gamma$ goes to zero, i.e., (see also \cite{PeRa2}),
\begin{equation}
\Gamma = \Gamma_{max}\,\,\,\, {\rm at}\,\,\,\, 
q = q_{max} < q_{crit},\,\,\,\,
{\rm and}\,\,\,\,\Gamma = 0\,\,\,\,{\rm at}\,\,\,\, q = q_{crit}.
\label{def1}
\end{equation}
The values of $q_{max}$ and $q_{crit}$ are obtained from, see Eq. 
(\ref{0disp1}),
\begin{equation}
{\partial \Gamma \over \partial q}\Big\vert_{q=q_{max}} = 0
\,\,\,\,\,\,\,\,{\rm and}\,\,\,\,\,\,\,
q_{crit}^2 = {|u_1|^2\over \kappa_s},
\,\,\,\, {\rm at}\,\,\,\,u_1^2 < 0.
\label{crit}
\end{equation}
Thus, the critical wave number $q_{crit}$ does not depend on
the viscosity. However, the presence of viscosity reduces the
instability, see also Fig. 1 below.

{\it (ii) Rare collision regime: $\tau \to \infty$.}\\
In this case, we have from Eqs. (\ref{0disp}), (\ref{kappa})
and (\ref{gamma})
\begin{equation}
\Gamma^2 = |u_1|^2\,q^2 - \kappa_v^\prime \,q^2 -\kappa_s\,q^4,
\label{0disp3}
\end{equation}
where
\begin{equation}
\kappa_v^\prime = {4\over 3\,m}{P_{eq}\over \rho_{eq}}.
\label{0kappa1}
\end{equation}

The critical value $q_{crit}$ and the value $q_{max}$ are given by
\begin{equation}
q_{crit}^2 = {{|u_1|^2 - \kappa_v^\prime}\over \kappa_s},
\,\,\,\,\,\,\,\,\,\,\,\, q_{max}^2 = {1\over 2}\,q_{crit}^2.
\label{crit1}
\end{equation}
Thus, the distortion of the Fermi-surface leads to a decrease of the 
critical value $q_{crit}$, i.e., the Fermi-liquid drop becomes more 
stable with respect to the volume density fluctuations due to the 
dynamic Fermi-surface distortion effects.

\section{Numerical results and discussion}

In Fig. 1 we have plotted the instability growth rate $\Gamma$ as 
obtained from Eq. (\ref{0disp}). The calculation was performed for the 
Skyrme force SIII. The relaxation time was taken in the form
$\tau = \hbar\,\alpha/T^2$ \cite{AbKh} with $\alpha = 9.2\,$MeV and 
$\alpha = 2.6\,$MeV \cite{KoPlSh1} and the bulk density $\rho_0$ was 
taken as $\rho_0 = 0.3\,\rho_{sat}$, where $\rho_{sat}$ is the saturated
density $\rho_{sat} = 0.1453\,fm^{-3}$. We show also the result for the 
nonviscous infinite nuclear matter and the nonviscous finite liquid drop 
neglecting Fermi surface distortion effects. In a finite system, the 
non-monotony behaviour of the instability growth rate as a function of 
the wave number $q$ is due to the anomalous dispersion term in Eq. 
(\ref{disp1}) created by the gradient terms in the equation of state.
We point out that the finite system becomes more stable with respect to 
short-wave-length density fluctuations at $q > q_{max}$. We can also see 
that the presence of viscosity decreases the instability. The strong 
decrease of instability in a Fermi liquid drop (FLD), when compared with
the corresponding result for the usual liquid drop (LD), is because of 
the Fermi surface distortion effects. In Fig. 2, this peculiarity of the 
FLD can be seen in a transparent way for both the infinite nuclear matter 
and the finite Fermi liquid drop.

For a saturated nuclear liquid one has for the force parameters 
$t_0 < 0,\,\,\, t_3 > 0$ and $t_s > 0$. Thus, the critical value 
$q_{crit}$, Eq. (\ref{crit}), increases with decreasing bulk density 
$\rho_0$ at $u_1^2 < 0$, see also Eq. (\ref{Fl}). The existence of the 
critical wave number $q_{crit}$ for an unstable mode is a feature of 
the finite system. The growth rate $\Gamma$ depends on the multipolarity 
$L$ of the nuclear density distortion and on the position of the 
eigenvalue, $q_L$, in the interval of $q = 0\,\div\,q_{crit}$ 
\cite{PeRa2}. For a given $R_0$, the value of $q_L$ increases with $L$ 
for $L \geq 2$ because of the boundary condition (\ref{sec1}), see 
Table 1. That means that if $q_L < q_{max}$ the instability increases 
with $L$ and the nucleus becomes more unstable with respect to an 
internal clusterization to small pieces (high multipolarity regime)
rather than to binary fission (low multipolarity regime). In contrast, 
the binary fission is preferable if $q_{max} < q_L < q_{crit}$. 

We give in Table 1 the values $q_L/k_F$ for two nuclei, $^{208}Pb$ 
and $^{40}Ca$ as obtained from Eq. (\ref{sec1}). The calculations 
were performed with the surface tension parameter 
$4\,\pi\,r_0^2\,\sigma = 17.2\,$MeV. We point out that the value of
$q_{max}$ is given here by $q_{max}/k_F = 0.69$. In Fig. 3 we have 
plotted the instability growth rate at $T=6\,$MeV and 
$\alpha = 9.2\,$MeV as function of the multipolarity $L$ of the 
particle density fluctuations for two nuclei $^{208}Pb$ and $^{40}Ca$.
As is seen from Fig. 3, the lowest values of $L \leq 3$ give the
contribution to the instability growth rate $\Gamma$ for the nucleus 
$^{40}Ca$. Thus, the nucleus $^{40}Ca$ is unstable with respect to 
the fission under the conditions considered above. In contrast, the 
instability growth rate of the nucleus $^{208}Pb$ includes the higher 
multipolarity $L \leq 8$ and this nucleus has to be unstable with 
respect to multifragmentation.

\section{Summary and conclusion}

Starting from the fluid dynamic equation of motion for the Fermi liquid 
drop with a sharp surface, we have derived the dispersion relations 
(\ref{disp1}) and (\ref{0disp}) for both the stable and the unstable 
regime. The dispersion relations are influenced strongly by the 
Fermi-surface distortion effect and the anomalous dispersion caused by 
the finiteness of the system. The presence of the Fermi surface distortion 
enhances the stiffness coefficient for a stable mode and reduces the
instability growth rate for an unstable one.

We have shown that the instability growth rate in an unstable finite
system is a non-monotony function of the wave number $q$ because of the 
anomalous dispersion term. This is in contrast with the infinite nuclear 
matter case where the instability growth rate increases with $q$. The 
non-monotony behaviour of the instability growth rate $\Gamma (q)$ in 
a finite Fermi liquid drop is accompanied with two characteristic wave 
numbers $q_{max}$ and $q_{crit}$, see Eqs. (\ref{def1}) and (\ref{crit}). 
The distortion of the Fermi-surface leads to a decrease of the critical 
value $q_{crit}$. The decay mode of an unstable Fermi liquid drop depends 
on the location of the eigen wave number $q$ on the slope of the curve 
$\Gamma (q)$. The Fermi liquid drop is more unstable with respect to
multifragmentation if $q < q_{max}$ and the binary fission is preferable 
if $q > q_{max}$. This is because the eigen wave number $q_L$, derived 
from the secular equation (\ref{sec1}), increases with the multipolarity 
$L$ of the particle density fluctuations. As an example, we have 
demonstrated this phenomenon in the case of hot nuclei $^{40}Ca$ and 
$^{208}Pb$. The nucleus $^{40}Ca$ is more unstable with respect to the 
short wave fluctuations and prefers to decay into the binary fission 
channel. The multifragmentation channel is preferable at the development
of the instability in the heavy nucleus $^{208}Pb$.

\section{Acknowledgements}
This work was supported in part by the US Department of Energy under 
grant \# DOE-FG03-93ER40773 and the INTAS under grant \# 93-0151. We are 
grateful for this financial support. One of us (V.M.K.) thanks the 
Cyclotron Institute at Texas A\&M University for the kind hospitality.

\newpage

\begin{table}
\caption{Values of $q_L/k_F$, as obtained from Eq. (16) for 
the surface tension parameter $\sigma$ derived from 
$4\,\pi\,r_0^2\,\sigma = 17.2\,$MeV and $T=1\,$MeV, for two nuclei 
$^{208}Pb$ and $^{40}Ca$.}

\bigskip
\begin{tabular}{cccccccc}
     L      &  2   &  3   & 4    & 5    &  6   & 7    &  8   \\
\tableline
$^{40}Ca$   & 0.986 & 1.225 & 1.457 & 1.684 & 1.907 & 2.128 & 2.347   \\
\hline
$^{208}Pb$  & 0.569 & 0.707 & 0.841 & 0.972 & 1.101 & 1.228 & 1.355   \\
\end{tabular}
\end{table}

\newpage

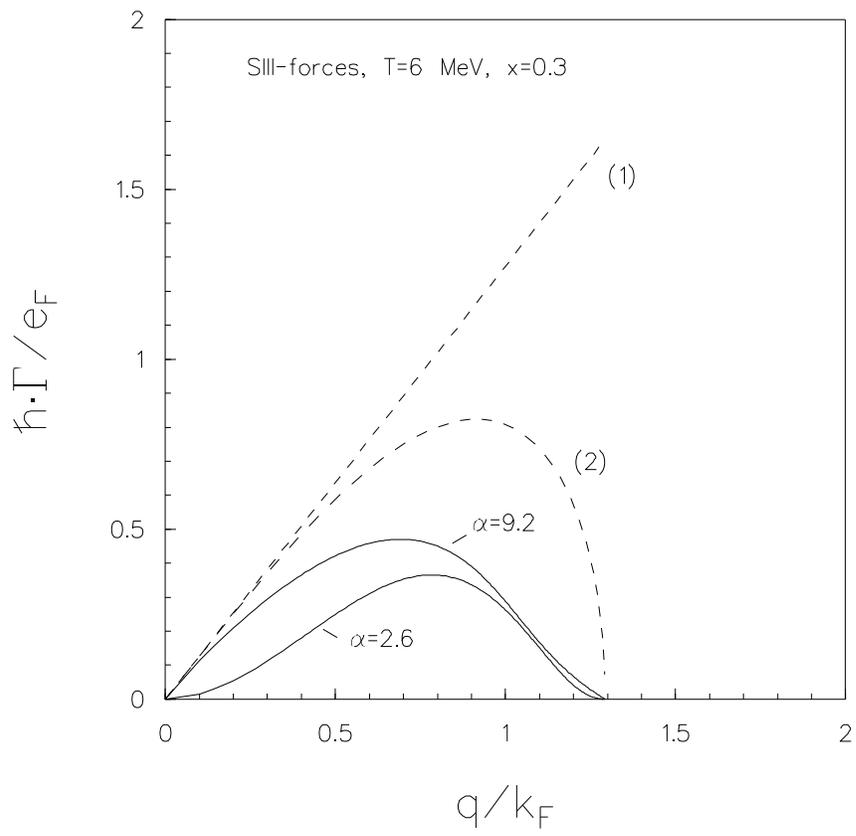
\begin{figure}
\caption{The dependence of the instability growth rate $\Gamma$ on the
wave number $q$. The calculations were performed for Skyrme force SIII,
temperature $T=6\,$MeV and density $\rho_0 = x\,\rho_{sat}$ with $x=0.3$. 
The solid curves are for the Fermi liquid drop from Eq. (17); the values
$\alpha = 9.2\,$MeV and $\alpha = 2.6\,$MeV of the relaxation time 
parameters are shown as labels to the curves. The dashed lines are the 
results for the nonviscous liquid without the Fermi surface distortion 
effects: curve (1) is the result for an infinite matter and curve (2) is 
for a finite liquid drop.}
\end{figure}

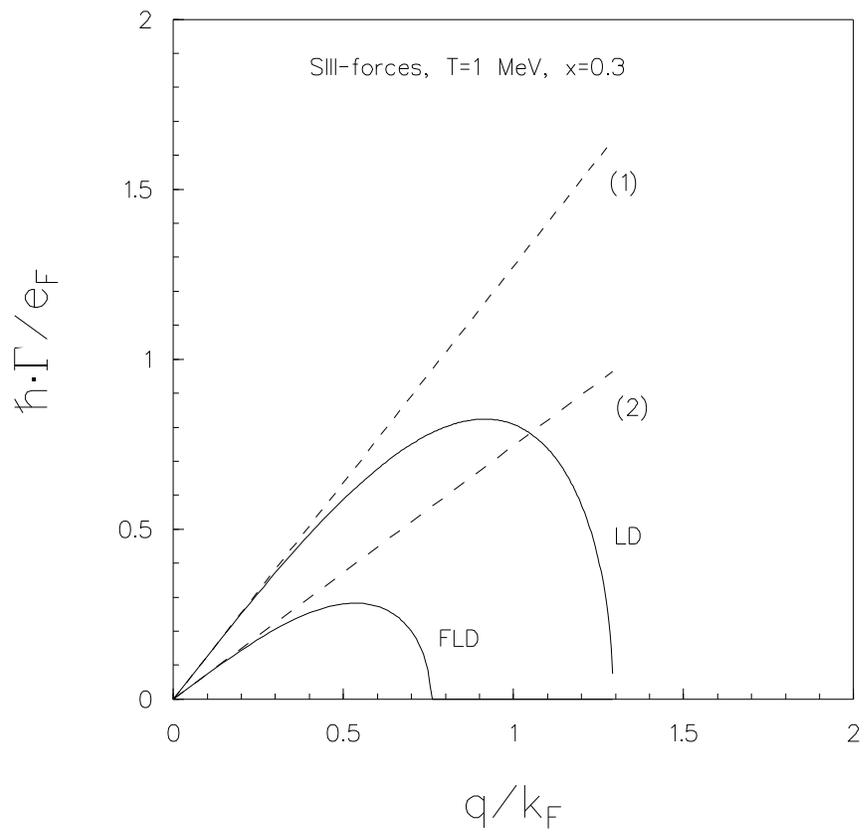
\begin{figure}
\caption{The dependence of the instability growth rate $\Gamma$ on the
wave number $q$ for the infinite matter (dashed lines) and the finite
system. The curves FLD and (2) are for the Fermi liquid; the curves LD 
and (1) are for the usual liquid, i.e., neglecting the Fermi surface 
distortion effects. All calculations were performed with $\alpha = 0$ 
and $T=1\,$MeV and with the force parameters and $\rho_0$ as in Fig. 1.}
\end{figure}

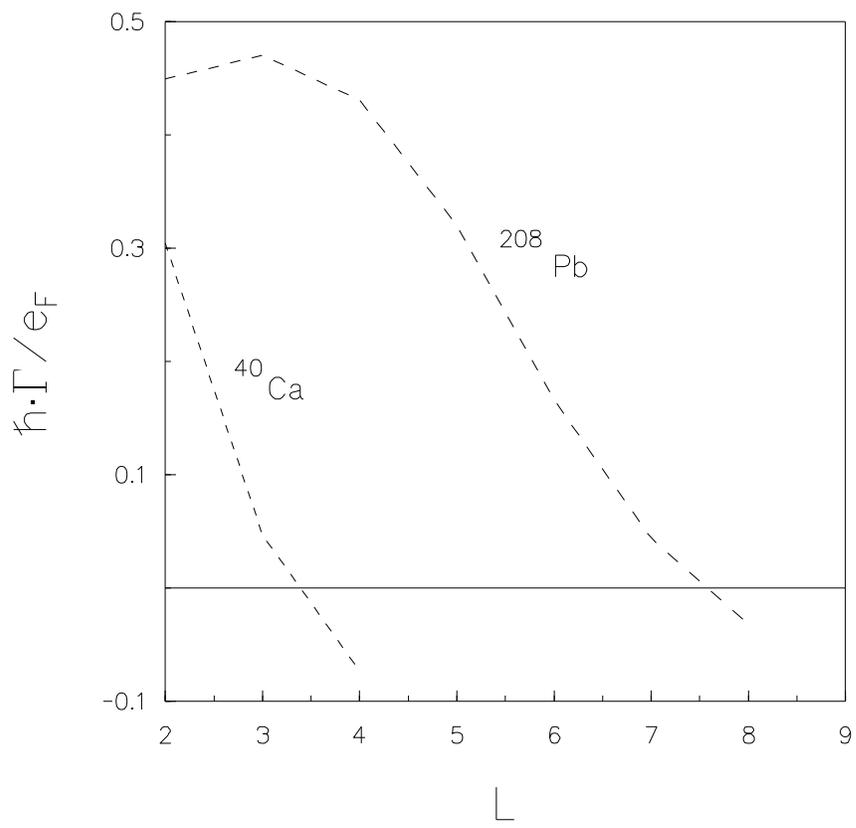
\begin{figure}
\caption{The dependence of the instability growth rate $\Gamma$ on the
multipolarity $L$ of the particle density fluctuations for two nuclei 
$^{208}Pb$ and $^{40}Ca$. The calculations were performed using the FLD 
results of Fig. 1 at $\alpha = 9.2\,$MeV and $T=1\,$MeV and the surface 
tension parameter $\sigma$ derived as in Table 1.}
\end{figure}

\end{document}